\documentclass{ws-procs10x7}

\usepackage{graphicx}

\begin{document}

\title{Review of Solar and Reactor Neutrinos}

\author{A.~W.~P.~Poon }
\address{Institute for Nuclear and Particle Astrophysics, Nuclear Science Divsion \\ 
Lawrence Berkeley National Laboratory \\ 1 Cyclotron Road, Berkeley, CA 94720, USA \\ 
E-mail: awpoon@lbl.gov}  

\twocolumn[\maketitle\abstract{Over the last several years, experiments
have conclusively demonstrated that neutrinos are massive and that they
mix.  There is now direct evidence for $\nu_e$s from the Sun
transforming into other active flavors while en route to the Earth. 
The disappearance of reactor $\bar{\nu}_e$s, predicted under the
assumption of neutrino oscillation, has also been observed. In this
paper, recent results from solar and reactor neutrino experiments and
their implications are reviewed.  In addition, some of the future
experimental endeavors in solar and reactor neutrinos are presented.}]

\section{Introduction}

From the 1960s to just a few years ago, solar neutrino experiments had
been observing fewer neutrinos than what were predicted by detail
models of the
Sun\cite{bib:bp01,bib:bp04,bib:bsb04,bib:tc99,bib:tc01,bib:tc04}.  The
radiochemical experiments, which used $^{37}$Cl\cite{bib:homestake98} 
and $^{71}$Ga\cite{bib:sage02,bib:gallex99,bib:gno02} as targets, were
sensitive exclusively to $\nu_e$.  The real-time water Cherenkov
detector Super-Kamiokande\cite{bib:skdet,bib:sk01a,bib:sk01b,bib:sk04}
(and its predecessor Kamiokande\cite{bib:kam96}) observes solar
neutrinos by $\nu$-e elastic scattering, and has sensitivity to all
active neutrino flavors.  However, its sensitivity to $\nu_\mu$ and
$\nu_\tau$ is only 1/6 of that for $\nu_e$, and the flavor content of
the observed solar neutrino events cannot be determined.

As these terrestrial detectors have different kinematic thresholds,
they probed different parts of the solar neutrino energy spectrum.  The
measured solar neutrino flux exhibited an energy dependence.  These
observations of an energy dependent flux deficit can be explained only
if the solar models are incomplete or neutrinos undergo flavor
transformation while in transit to the Earth. 
Table~\ref{tbl:solarnuexp} shows a comparison of the predicted and the
observed solar neutrino fluxes for these experiments.

Since 2001, significant advances have been made in solar neutrino
physics.  The Sudbury Neutrino Observatory
(SNO)\cite{bib:snodet,bib:snocc,bib:snonc,bib:snodn,bib:saltprl,bib:saltprc}  
has conclusively demonstrated that a significant fraction of
$\nu_e$s that are produced in the solar core transforms into other
active flavors.  One of the most favored explanation for this flavor
transformation is matter-enhanced neutrino oscillation, or the
Mikheyev-Smirnov-Wolfenstein (MSW) effect\cite{bib:msw}.  The KamLAND
experiment\cite{bib:kldet,bib:klrate,bib:klspec} observes the
disappearance of reactor $\bar{\nu}_e$s that is predicted from the
neutrino mixing parameters derived from global MSW analyses of solar
neutrino results.  This provides very strong evidence that MSW
oscillation is the underlying mechanism in solar neutrino flavor
transformation.  In this paper, these advances in solar and reactor
neutrino experiments and their physical implications are discussed.  A
brief overview of the future program in solar neutrinos and reactor
anti-neutrinos will also be presented.

\begin{table*}[t]
    \caption{\label{tbl:solarnuexp}Summary of solar neutrino
    observations at different terrestrial detectors before 2002.  The
    Bahcall-Pinsonneault (BP2001) model predictions of the solar
    neutrino flux are presented in this table.  The experimental values
    are shown with statistical uncertainties listed first, followed by
    systematic uncertainties.  The Solar Neutrino Unit (SNU) is a
    measure of solar neutrino interaction rate, and is defined as 1
    interaction per 10$^{-36}$ target atom per second.  For the
    Kamiokande and Super-Kamiokande experiments, the predicted and
    measured neutrino fluxes are listed in units of
    $10^{6}$~cm$^{-2}$~s$^{-1}$.}
   \begin{center}
    \begin{tabular}{llll} \hline
	Experiment & Measured Rate/Flux & SSM Prediction (BP2001)\cite{bib:bp01} \\ \hline
	
	Homestake\cite{bib:homestake98}  ($^{37}$Cl)  &
	2.56 $\pm$ 0.16 $\pm$ 0.16 SNU & 7.6 $^{+1.3}_{-1.1}$ SNU 
	 \\ \hline
	
	SAGE\cite{bib:sage02} ($^{71}$Ga) &
	70.8 $^{+5.3}_{-5.2}$ $^{+3.7}_{-3.2}$ SNU & 
	\\
	
	Gallex\cite{bib:gallex99} ($^{71}$Ga) &
	77.5 $\pm$ 6.2 $^{+4.3}_{-4.7}$ SNU & 128 $^{+9}_{-7}$ SNU 
	\\ 
	GNO\cite{bib:gno02} ($^{71}$Ga) &
	62.9 $^{+5.5}_{-5.3}$ $\pm$ 2.5 SNU & 
	\\ \hline

	Kamiokande\cite{bib:kam96} ($\nu e$) &	
	2.80 $\pm$ 0.19 $\pm$ 0.33 & 
	5.05 $\left(1^{+0.20}_{-0.16}\right)$ \\ 
	
	Super-Kamiokande\cite{bib:sk01a} ($\nu e$) &	
	2.32 $\pm$ 0.03 $^{+0.08}_{-0.07}$ & 
	 \\ \hline

    \end{tabular}
    \end{center}
\end{table*}
 
\section{Solar Neutrino Flux Measurements at Super-Kamiokande}

The Super-Kamiokande (SK) detector is a 50000-ton water Cherenkov
detector located in the Kamioka mine, Gifu prefecture, Japan. During
the first phase of the experiment SK-I (April 1996 to July 2001),
approximately 11200 20-inch-diameter photomultiplier tubes (PMTs) were
mounted on a cylindrical tank to detect Cherenkov light from neutrino
interactions in the inner detector.  Since December 2002, the
experiment has been operating in its second phase (SK-II) with
approximately 5200 PMTs in its inner detector.  An additional 1885
8-inch-diameter PMTs are used as a cosmic veto.

\subsection{Super-Kamiokande-I}
\begin{figure}
\begin{center}
\includegraphics[width=190pt]{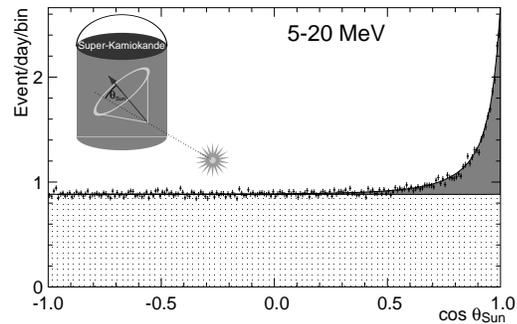}
\caption{\label{fig:sk1-soldir}  Angular distribution of solar neutrino
event candidates in the 1496-day SK-I data set.  The shaded area is the
solar neutrino elastic scattering peak, and the dotted area represents
backgrounds in the candidate data set.}
\end{center}
\end{figure}

In SK-I and SK-II, neutrinos from the Sun are detected through the
elastic scattering process $\nu e \rightarrow \nu e$.  Because of the
strong directionality in this process, the reconstructed direction of
the scattered electron is strongly correlated to the direction of the
incident neutrinos.  The sharp elastic scattering peak in the angular
distribution for events with a total electron energy of 5$<E<$20~MeV in
the SK-I  data set is shown in Figure~\ref{fig:sk1-soldir}.  This data
set spans 1496 days (May 31, 1996 to July 15, 2001), and the solar
neutrino flux is extracted by statistically separating the solar
neutrino signal and the backgrounds using this angular distribution. 
At the analysis threshold of $E$=5~MeV, the primary signal is $\nu_e$s
from $^8$B decays in the solar interior.  The extracted solar $^8$B
neutrino flux in this SK-I data set ($\phi^{\rm ES}_{\rm SK-I}$)
is\cite{bib:sknote} (in units of $10^6\, \mbox{cm}^{-2}\mbox{s}^{-1}$):
\begin{displaymath}
    \phi^{\rm ES}_{\rm SK-I} = 2.35 \, \pm \, 0.02 (\mbox{stat.}) \, \pm \, 0.08 (\mbox{sys.})
\end{displaymath}
When comparing this measured flux to the BP2001\cite{bib:bp01} and
BP2004\cite{bib:bp04} model predictions:
\begin{eqnarray}
    \frac{\phi^{\rm ES}_{\rm SK-I}}{\phi_{\rm BP2001}} &=& 0.465 \, \pm \, 0.005 (\mbox{stat.}) \, ^{+0.016}_{-0.015} (\mbox{sys.}) \nonumber \\
    \frac{\phi^{\rm ES}_{\rm SK-I}}{\phi_{\rm BP2004}} &=& 0.406 \, \pm \, 0.004 (\mbox{stat.}) \, ^{+0.014}_{-0.013} (\mbox{sys.}), \nonumber
\end{eqnarray}
where the model uncertainties ($\sim$20\%) have not been included in
the systematic uncertainties above.

\subsection{Super-Kamiokande-II}

\begin{figure}
\begin{center}
\includegraphics[width=190pt]{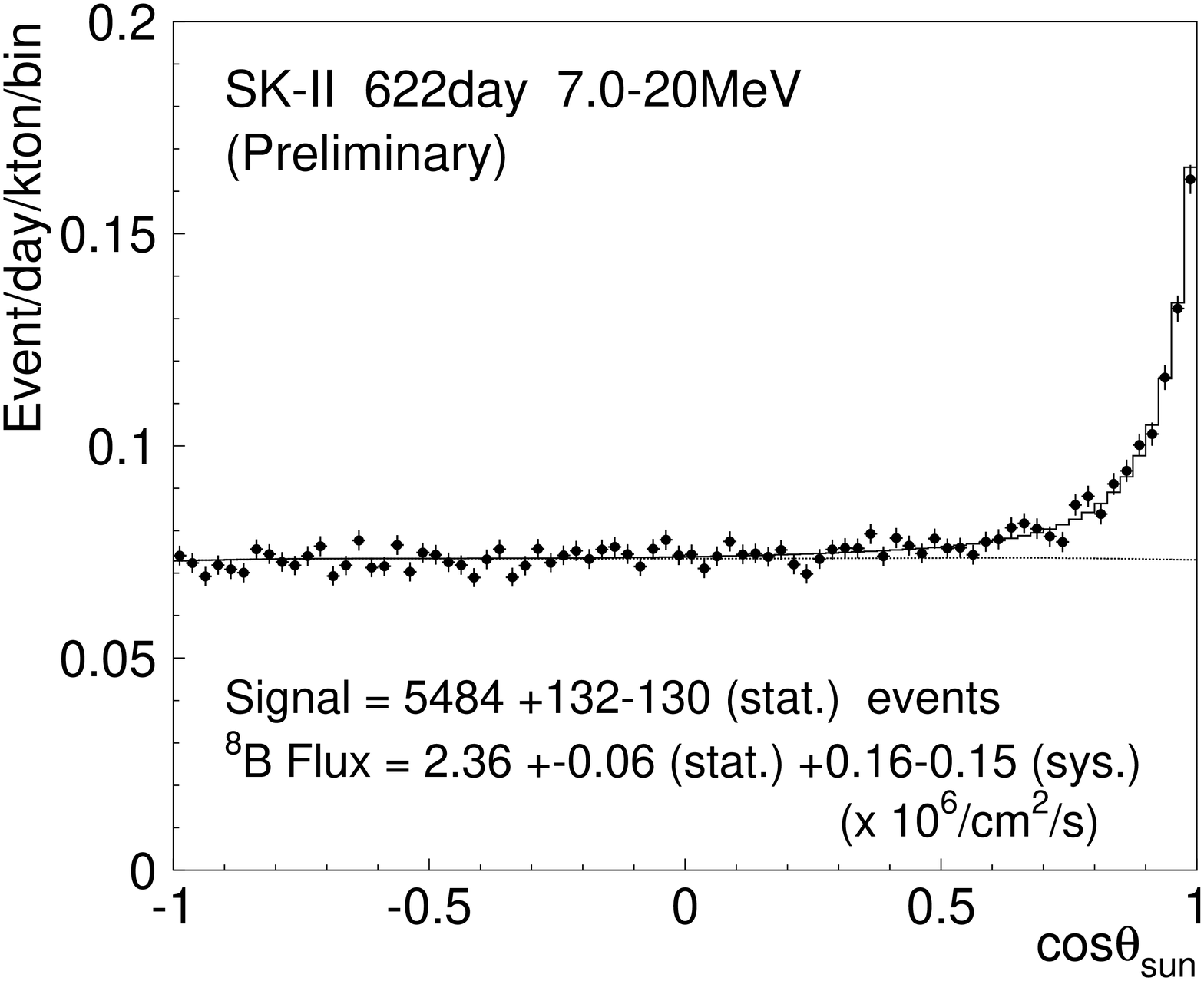}
\caption{\label{fig:sk2-soldir-all}  Angular distribution of solar
neutrino event candidates in a 622-day data set in SK-II.  For the
first 159 days of these data, the energy threshold for the analysis was
set at $E=$8~MeV, and it was lowered to 7~MeV for the remaining 463
days of data.}
\end{center}
\end{figure}

With only about half of the photocathode coverage as SK-I, significant
improvements have been made to the trigger system in the SK-II detector
in order to maintain high trigger efficiency for solar neutrino events.
 The improved trigger system can trigger with 100\% efficiency at
$E\sim$6.5~MeV.    Results from a 622-day SK-II data set have recently
been released.  The solar angular distribution plot is shown in
Figure~\ref{fig:sk2-soldir-all}.  For the first 159 days (Dec. 24, 2002
to July 15, 2003) of these data, the energy threshold for the analysis
was set at $E=$8~MeV, and it was lowered to 7~MeV for the remaining 463
days (Jul. 15, 2003 to Mar. 19, 2005).  The extracted solar $^8$B
neutrino flux in this 622-day SK-II data set ($\phi^{\rm ES}_{\rm
SK-II}$) is (in units of $10^6\, \mbox{cm}^{-2}\mbox{s}^{-1}$):
\begin{displaymath}
    \phi^{\rm ES}_{\rm SK-II} = 2.36 \, \pm \, 0.06 (\mbox{stat.}) \, 
        ^{+0.16}_{-0.15} (\mbox{sys.}),
\end{displaymath}
which is consistent with the SK-I results.

\section{Sudbury Neutrino Observatory}

The Sudbury Neutrino Observatory (SNO) detector is a 1000-tonne heavy
water (D$_2$O) Cherenkov detector located near Sudbury, Ontario,
Canada.  Approximately 9500 8-inch-diameter PMTs are mounted on a
spherical geodesic structure to detect Cherenkov light resulting from
neutrino interactions.   It can make simultaneous measurements of the
$\nu_e$ flux from $^{8}$B decay in the Sun and the flux of all active
neutrino flavors\cite{bib:chen} through the following reactions:
\[\begin{array}{lcll}
    \nu_{e}+d & \rightarrow & p+p+e^{-} & \hspace{0.5in} \mbox{(CC)}\\ 
    \nu_{x}+d & \rightarrow & p+n+\nu_{x} & \hspace{0.5in} \mbox{(NC)} \\
    \nu_{x}+e^{-} & \rightarrow &  \nu_{x}+e^{-} & \hspace{0.5in} \mbox{(ES)} \\
\end{array}\]
The charged-current (CC) reaction on the deuteron is sensitive
exclusively to $\nu_e$, and the neutral-current (NC) reaction has equal
sensitivity to all active neutrino flavors ($\nu_x$, $x=e,\mu,\tau$). 
Similar to the Super-Kamiokande experiment, elastic scattering (ES) on
electron is also sensitive to all active flavors, but with reduced
sensitivity to $\nu_\mu$ and $\nu_\tau$.  If the measured total $\nu_x$
flux (through the NC channel) is greater than the measured $\nu_e$ flux
(through the CC channel), it would conclusively demonstrate that solar
$\nu_e$s have undergone flavor transformation since their production in
the solar core.  Alternatively, this flavor transformation can be
demonstrated by comparing the $\nu_x$ flux deduced from the ES channel
to the $\nu_e$ flux.  

\subsection{Pure D$_2$O phase}

The first phase of the SNO experiment (SNO-I) used a pure D$_2$O
target.  The free neutron from the NC interaction is thermalized, and
in 30\% of the time, a 6.25-MeV $\gamma$ ray is emitted following the
neutron capture by deuteron.   In 2001, the SNO collaboration published
a measurement of the $\nu_e$ flux, based on a 241-day data set taken
from Nov. 2, 1999 to Jan 15, 2001.  At an electron kinetic energy
$T_{\rm eff}$ threshold of 6.75~MeV\cite{bib:snocc}, the measured
$\nu_e$ and $\nu_x$ fluxes through the CC $\phi^{\rm CC}_{\rm SNO-I}$
and ES $\phi^{\rm ES}_{\rm SNO-I}$ channels are (in units of $10^6\,
\mbox{cm}^{-2}\mbox{s}^{-1}$):
\begin{eqnarray}
    \phi^{\rm CC}_{\rm SNO-I} &=& 
    1.75\pm 0.07\mbox{(stat.)}^{+0.12}_{-0.11}\mbox{(sys.)} \nonumber \\ 
    \phi^{\rm ES}_{\rm SNO-I} &=& 
    2.39\pm 0.34\mbox{(stat.)}^{+0.16}_{-0.14}\mbox{(sys.)}. \nonumber 
\end{eqnarray}
The measured $\phi^{\rm ES}_{\rm SNO-I}$ agrees with that from the SK-I
detector $\phi^{\rm ES}_{\rm SK-I}$.  But a comparison of $\phi^{\rm
CC}_{\rm SNO-I}$ to $\phi^{\rm ES}_{\rm SK-I}$, after adjusting for the
difference in the energy response of the two detectors, yields (in
units of $10^6\, \mbox{cm}^{-2}\mbox{s}^{-1}$)
\begin{displaymath}
   \phi^{\rm ES}_{\rm SK-I}(\nu_x) - \phi^{\rm ES}_{\rm SNO-I}(\nu_e) = 0.57 \pm 0.17,
\end{displaymath}
which is 3.3$\sigma$ away from 0.  This measurement not only confirmed
previous observations of the solar neutrino deficit from different
experiments, it also provided the first indirect evidence, when
combined with the SK-I results, that neutrino flavor transformation
might be the solution to this long-standing deficit.

In 2002, the SNO collaboration reported a measurement of the total
active neutrino flux through the NC channel\cite{bib:snonc}.  This
measurement used a $T_{\rm eff}$ threshold of 5 MeV and was based the
306-day data set (Nov. 2, 1999 to May 28, 2001).  Under the assumption
of an undistorted $^8$B $\nu_e$ spectrum, the non-$\nu_e$ component
($\phi^{\mu\tau}_{\rm SNO-I}$) of the total active neutrino flux is (in
units of $10^6\, \mbox{cm}^{-2}\mbox{s}^{-1}$)
\begin{displaymath}
    \phi^{\mu\tau}_{\rm SNO-I} =   3.41\pm 0.45\mbox{(stat.)}^{+0.48}_{-0.45}\mbox{(sys.)},
\end{displaymath}
which is 5.3$\sigma$ away from 0.  This result was the first direct
evidence that demonstrated neutrino flavor transformation.  The
measured total active neutrino flux confirmed the solar model
predictions and provided the definitive solution to the solar neutrino
deficit problem.

\subsection{Salt Phase}

\begin{figure}
\begin{center}
\includegraphics[width=190pt]{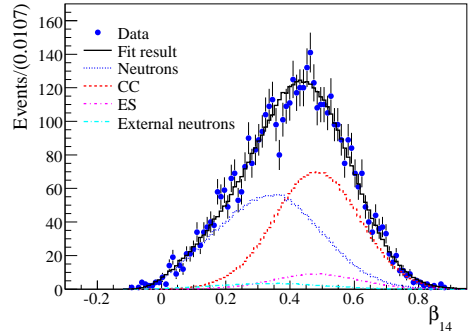}
\caption{\label{fig:b14_ucon}  Statistical separation of CC and NC
events using Cherenkov light isotropy in SNO-II.  The measure of
isotropy, $\beta_{14}$, is a function with Legendre polynomials of
order 1 and 4 as its bases.  Because there were multiple $\gamma$s in
the NC signal, the  Cherenkov light distribution was more diffuse
(smaller $\beta_{14}$). }
\end{center}
\end{figure}

In phase two of the SNO experiment (SNO-II), 2 tonnes of NaCl were
added to the D$_2$O target in order to enhance the detection efficiency
of the NC channel.  The free neutron from the NC channel was
thermalized in the D$_2$O and subsequently captured by a $^{35}$Cl
nucleus, which resulted in the emission of a $\gamma$-ray cascade with a
total energy of 8.6~MeV.  The neutron capture efficiency increased
three folds from SNO-I.  The CC signal involved a single electron and multiple
$\gamma$s were emitted in the NC channel.  This difference in the number of 
particles in the final state resulted in a difference in the isotropy of the Cherenkov 
light distribution.  The CC and the NC signals could be statistically separated by 
this isotropy difference.  This separation for events with $T_{\rm eff}>$5.5~MeV 
is shown in Fig.~\ref{fig:b14_ucon} for the 391-day data set (taken from Jul. 26,
2001 to Aug. 28, 2003).  This use of light isotropy also removed the
need to constrain the $^{8}$B $\nu_e$ energy spectrum, which can be
distorted if the neutrino flavor transformation process is energy
dependent, as in SNO-I.  The measured energy-unconstrained $\nu_e$ and
$\nu_x$ fluxes through the different channels
are\cite{bib:saltprl,bib:saltprc} (in units of $10^6\,
\mbox{cm}^{-2}\mbox{s}^{-1}$):
\begin{eqnarray}
	\phi^{\rm uncon. CC}_{\rm SNO-II} &=& 1.68 \pm 0.06\mbox{(stat.)} 
	    ^{+0.08}_{-0.09}\mbox{(sys.)} \nonumber \\
	\phi^{\rm uncon. ES}_{\rm SNO-II} &=& 2.35 \pm 0.22\mbox{(stat.)} 
	    ^{+0.15}_{-0.15}\mbox{(sys.)} \nonumber \\
	\phi^{\rm uncon. NC}_{\rm SNO-II} &=& 4.94 \pm 0.21\mbox{(stat.)} 
	    ^{+0.38}_{-0.34}\mbox{(sys.)}. \nonumber 
\end{eqnarray}
The ratio of the $\nu_e$ flux and the total active neutrino flux is of
physical significance (which will be discussed later), and is
\begin{displaymath}
	\frac{\phi^{\rm uncon. CC}_{\rm SNO-II}}{\phi^{\rm uncon. NC}_{\rm SNO-II} } = 
	0.340 \pm 0.023 \mbox{(stat.)} ^{+0.029}_{-0.031} \mbox{(sys.)}.
\end{displaymath}

\section{Search for MSW Signatures in Solar Neutrinos}

Recent results from SNO and Super-Kamiokande have conclusively
demonstrated that neutrino flavor transformation is the solution to the
solar neutrino deficit.  The most favored mechanism for this
transformation is the Mikheyev-Smirnov-Wolfenstein (MSW)
matter-enhanced neutrino oscillation\cite{bib:msw}.   MSW oscillation
can be a resonant effect as opposed to vacuum oscillation, which is
simply the projection of the time evolution of eigenstates in free
space.  A resonant conversion of $\nu_e$ to other active flavors is
possible in MSW oscillation if the ambient matter density matches the
resonant density.  Two distinct signatures of the MSW effect are
distortion of the neutrino energy spectrum and a day-night asymmetry in
the measured neutrino flux.  The former signature arises from the
energy dependence in neutrino oscillation.  When the Sun is below a
detector's horizon, some of the oscillated solar neutrinos may revert
back to $\nu_e$ while traversing the Earth's interior.  This $\nu_e$
re-generation effect  would give an asymmetry in the measured $\nu_e$
fluxes during the day and the night.

\begin{figure}[t]
\begin{center}
\includegraphics[width=190pt]{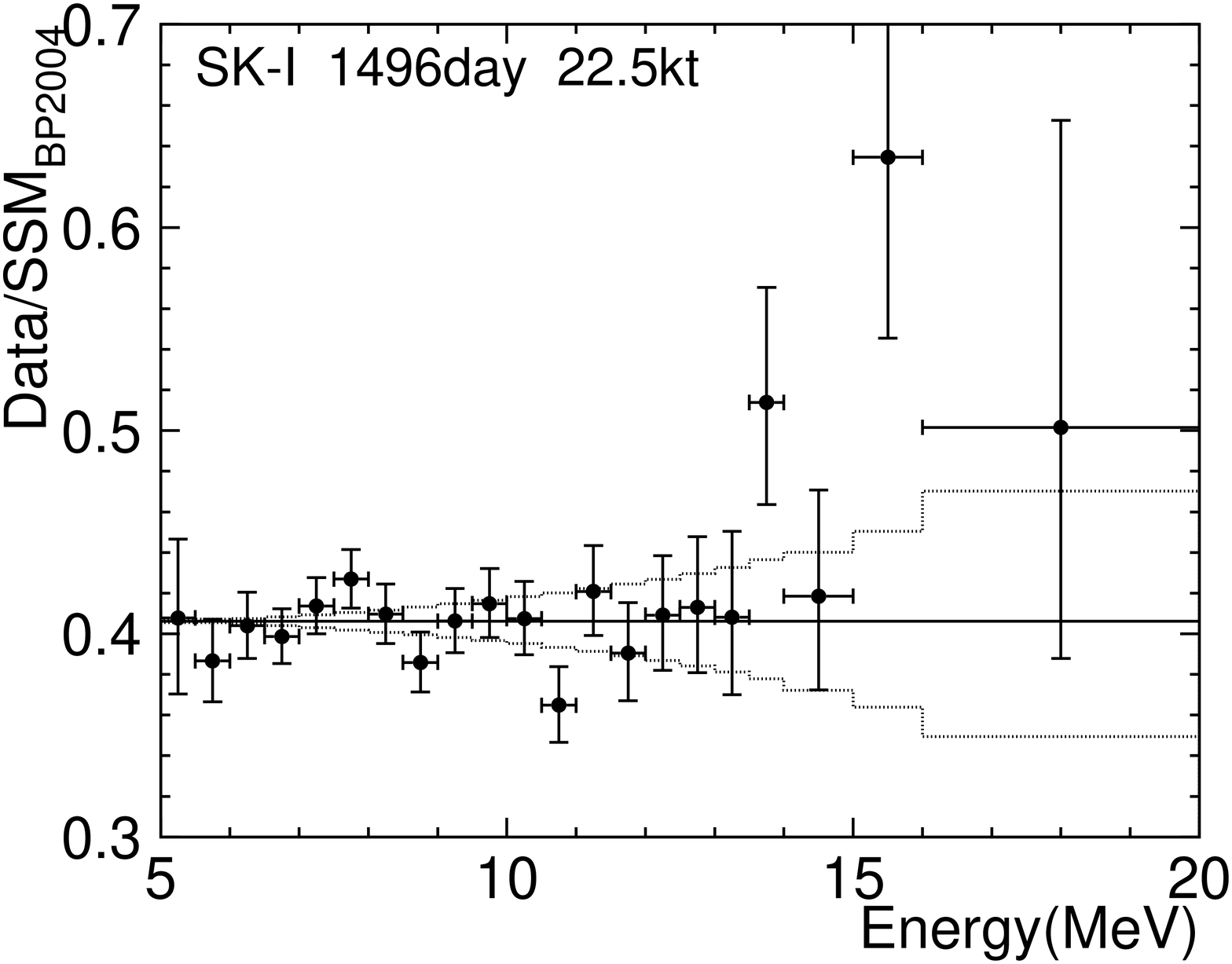}
\includegraphics[width=190pt]{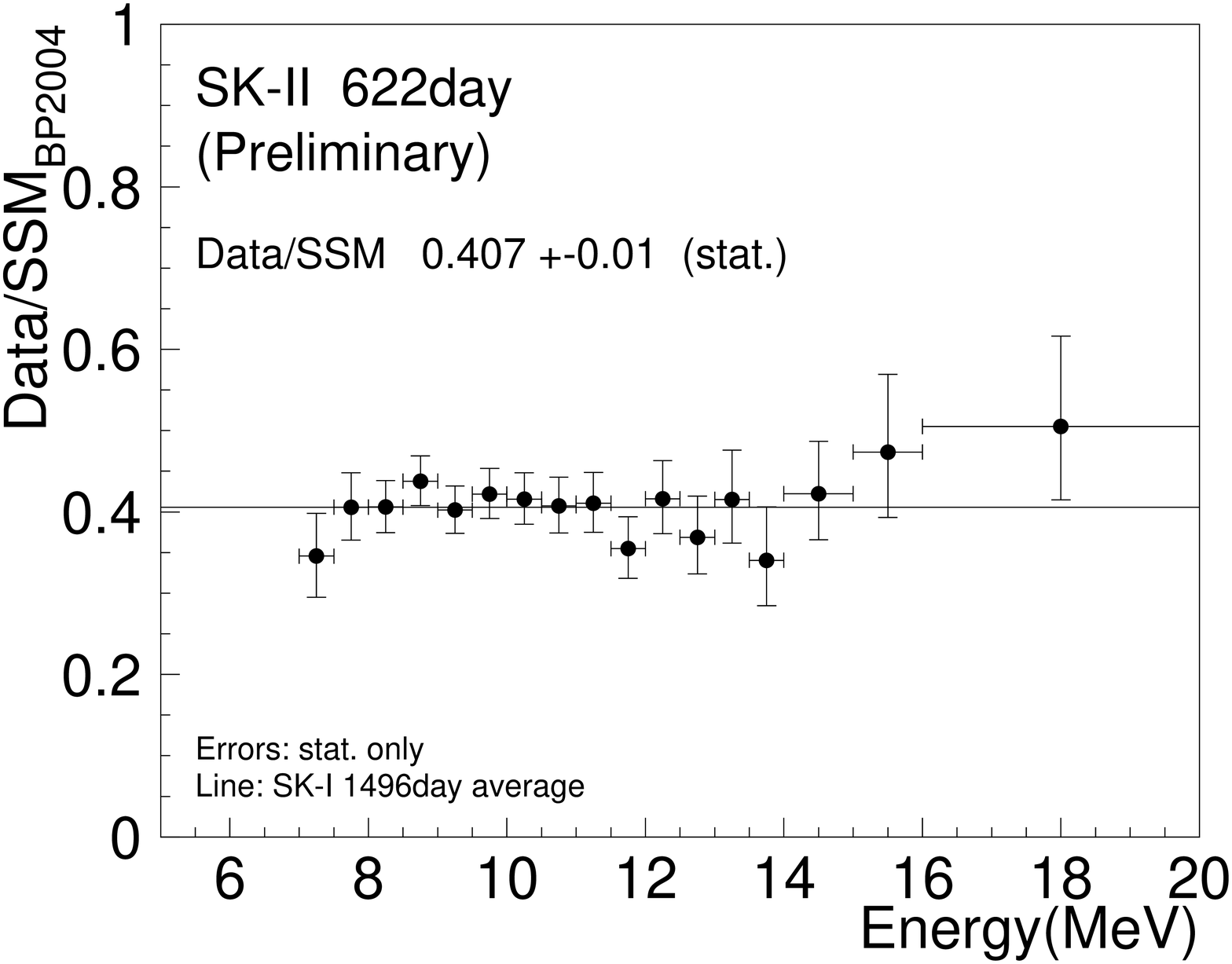}
\caption{\label{fig:skspect} Measured electron energy spectra from SK-I
and SK-II.  The measured spectra have been normalized to the BP2004
model predictions.  The solid lines indicate the mean of the measured
$^8$B neutrino flux.  The band in the SK-I spectrum represents the
energy-correlated systematic uncertainties in the measurement.}
\end{center}
\end{figure}

\begin{figure}
\begin{center}
\includegraphics[width=190pt]{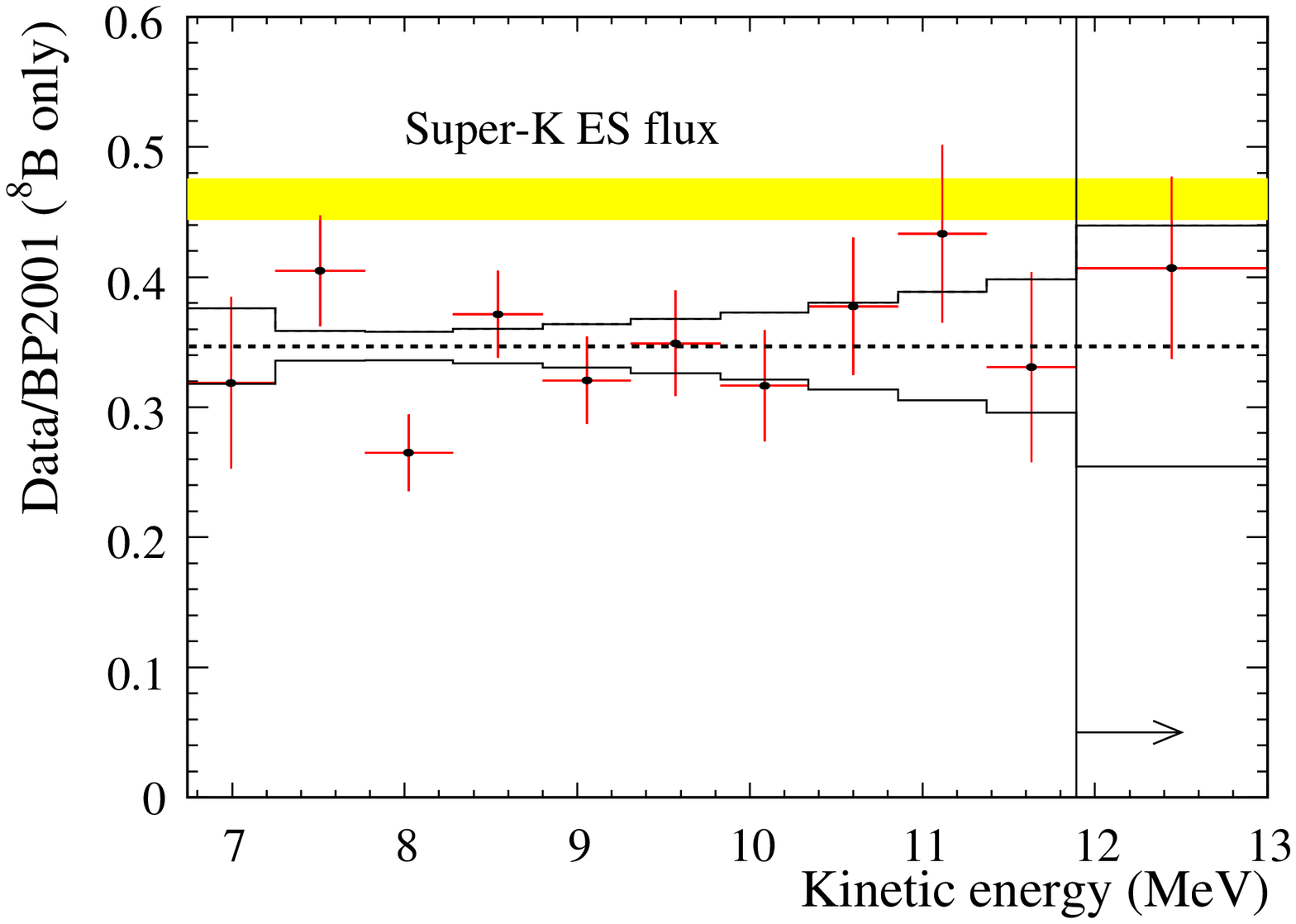}
\includegraphics[width=190pt]{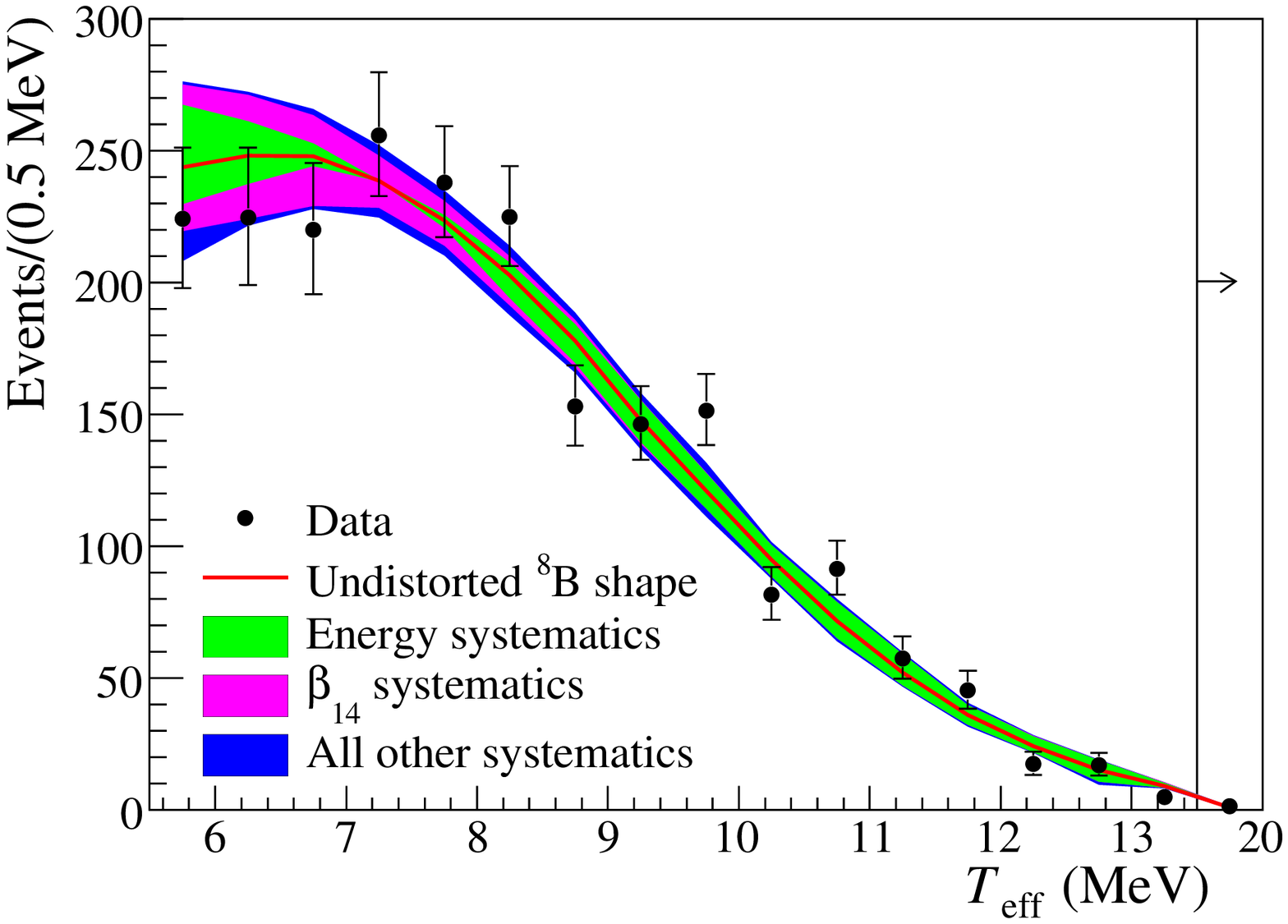}
\includegraphics[width=190pt]{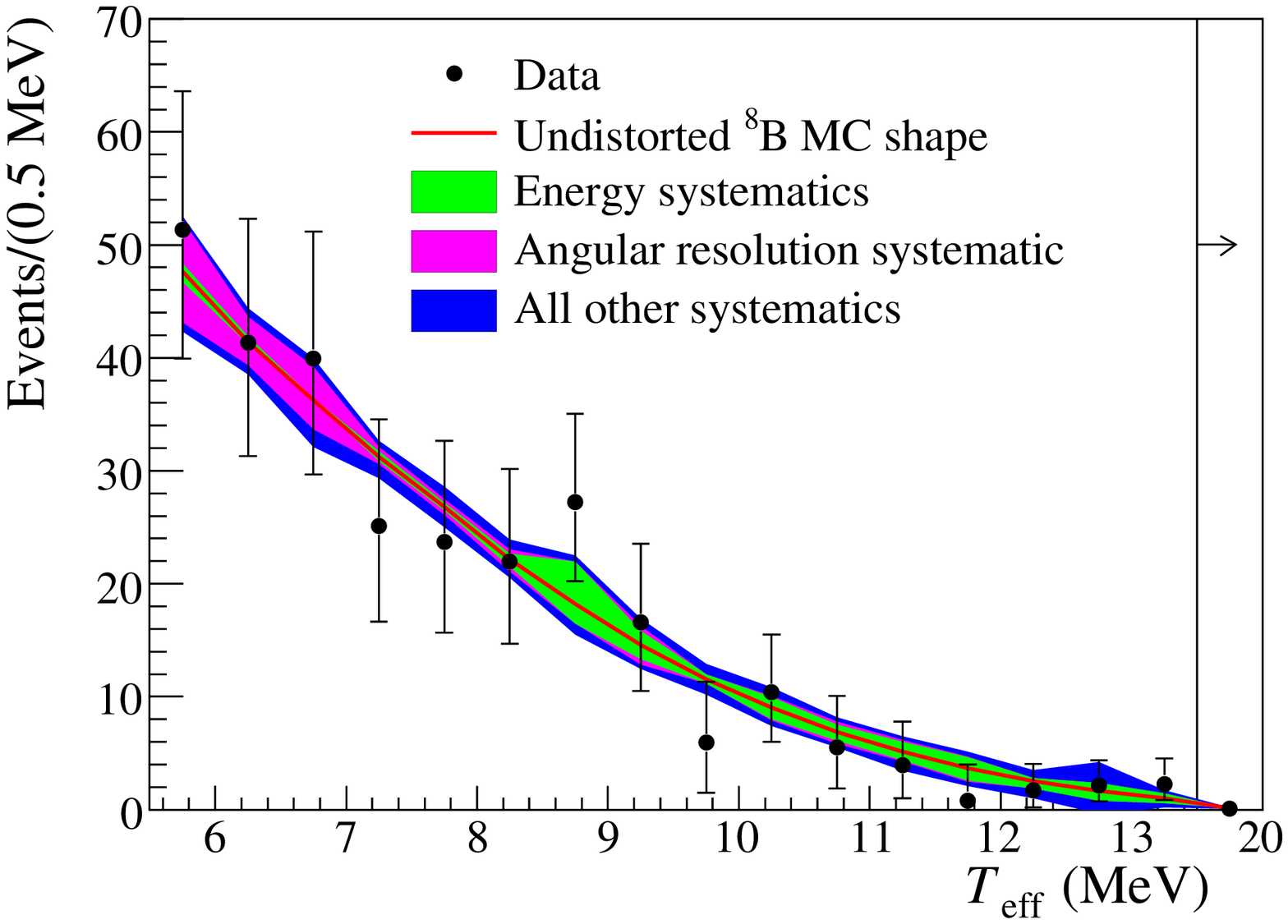}
\caption{\label{fig:snospect} Measured electron energy spectra from
SNO-I and SNO-II.  {\it Top}: The ratio of the measured SNO-I CC
electron kinetic energy spectrum to the expected undistorted kinetic
energy distribution for $^8$B neutrinos (in BP2001 model) with
correlated systematic uncertainties.  {\it Middle}:  The measured CC
electron kinetic energy spectrum in SNO-II is shown as the data points
(with statistical uncertainties only).  {\it Bottom}: The measured ES
electron kinetic energy spectrum in SNO-II is shown  as the data points
(with statistical uncertainties only).   In the last two plots, the
bands show the accumulated effect of different correlated systematic
uncertainties on the electron spectra expected from an undistorted
$^8$B neutrino spectrum.  }
\end{center}
\end{figure}

Both SNO and Super-Kamiokande have done extensive searches for these
two signatures in their data.  Figure~\ref{fig:skspect} shows the
measured electron spectra from SK-I and SK-II, whereas
Fig.~\ref{fig:snospect} shows the measured spectra from SNO-I and
SNO-II.  No statistically significant distortion is seen from either
experiment.

The Super-Kamiokande experiment defines the day-night asymmetry ratio
$A^{\rm SK}_{\rm DN}$ as\cite{bib:sk04}
\begin{displaymath}
   A^{\rm SK}_{\rm DN} = \frac{\Phi_{\rm D}-\Phi_{\rm N}}
                              {\frac{1}{2}(\Phi_{\rm D}+\Phi_{\rm N})},
\end{displaymath}
where $\Phi_{\rm D}$ is the measured neutrino flux when the Sun is
above the horizon, and $\Phi_{\rm N}$ is the corresponding flux when
the Sun is below the horizon.  The measured day-night asymmetries of
the solar neutrino flux by SK-I and SK-II are:
\begin{eqnarray}
	A^{\rm SK-I}_{\rm DN} &=& -0.021 \pm 0.020\mbox{(stat.)} 
	 ^{+0.013}_{-0.012} \mbox{(sys.)} \nonumber \\
	A^{\rm SK-II}_{\rm DN} &=& 0.014 \pm 0.049\mbox{(stat.)} 
	 ^{+0.024}_{-0.025} \mbox{(sys.)}. \nonumber
\end{eqnarray}
It should be noted that the flux measured by Super-Kamiokande is a
mixture of all three active neutrino flavors.

The SNO experiment has also measured the day-night asymmetry of the
measured neutrino flux\cite{bib:snodn,bib:saltprc}.  It should be noted
that the SNO and the Super-Kamiokande asymmetry ratios are defined
differently, such that  $A^{\rm SNO}_{\rm DN} = - A^{\rm SK}_{\rm DN}$.
 Because SNO can measure the $\nu_e$ flux and the total active neutrino
flux separately through the CC and the NC channels, it can determine
the day-night asymmetry for these fluxes separately.  In addition, the
asymmetry ratio can be determined with the day-night asymmetry in the
NC channel $A^{\rm SNO}_{\rm NC, DN}$ constrained to 0.  With the $^8$B
shape and $A^{\rm SNO}_{\rm NC, DN}=0$ constraints, the measured
day-night asymmetry in the $\nu_e$ flux in SNO-I and SNO-II are
\begin{eqnarray}
	A^{\rm SNO-I}_{e,{\rm DN}} &=& 0.021 \pm 0.049\mbox{(stat.)} 
	 ^{+0.013}_{-0.012} \mbox{(sys.)} \nonumber \\
	A^{\rm SNO-II}_{e,{\rm DN}} &=& -0.015 \pm 0.058\mbox{(stat.)} 
	 ^{+0.027}_{-0.027} \mbox{(sys.)}. \nonumber
\end{eqnarray}

Because of the presence of $\nu_\mu$ and $\nu_\tau$ in the
Super-Kamiokande measured flux, its day-night asymmetry is diluted by a
factor of 1.55\cite{bib:saltprc}. Assuming an energy-independent
conversion mechanism and only active neutrinos, the SK-I result scales
to a $\nu_e$ flux asymmetry $A_{e,{\rm SK-I}} = 0.033 \pm 0.031
^{+0.019}_{-0.020}$.  Combining the SNO-I and SNO-II values for
$A_{e,{\rm DN}}$ with the equivalent SK-I value ($A_{e,{\rm SK-I}}$)
gives $A_{e,{\rm combined}} = 0.035 \pm 0.027$.  No statistically
significant day-night asymmetry has been observed.

\section{MSW Interpretation of Solar Neutrino Data}

Although no direct evidence for the MSW effect has been observed, the
null hypothesis that no MSW oscillation in the solar neutrino results
is rejected at 5.6$\sigma$\cite{bib:fogli}.   There are two parameters
in a two-flavor, active neutrino oscillation model: $\Delta m^2$, which
is the difference of between the square of the eigenvalue of two
neutrino mass states; and $\tan^2\theta$, which quantifies the mixing
strength between the flavor and the mass eigenstates.  Each pair of
these parameters affects the total solar neutrino spectrum differently,
which can give rise to the energy dependence in the ratio between the
observed and the predicted neutrino fluxes in different detectors.   
Using the measured rates in the radiochemical ($^{37}$Cl and $^{71}$Ga)
experiments, the solar zenith angle distribution from the
Super-Kamiokande experiment, and the day and night energy spectra from
the SNO experiment, a global statistical analysis can then be performed
to determine the ($\Delta m^2$, $\tan^2\theta$) pair that best
describes the data\cite{bib:global_anl}.  The best-fit
parameters\cite{bib:saltprc} are found in the so-called ``Large Mixing
Angle'' (LMA) region:
\begin{eqnarray}
	\Delta m^2 &=& 6.5 ^{+4.4}_{-2.3} \times 10^{-5} \mbox{eV}^2 \nonumber \\
	\tan^2\theta &=& 0.45 ^{+0.09}_{-0.08}. \nonumber
\end{eqnarray}

There are two implications to these results.  First, maximal mixing
(i.e. $\tan^2\theta = 1$) is ruled out at very high significance.  This
is in contrary to the atmospheric neutrino sector, where maximal mixing
is the preferred scenario.  Second, because the ``dark-side''
($\tan^2\theta > 1$) is also ruled out, a mass hierarchy of $m_2 > m_1$
is implied.  

\begin{figure}
\begin{center}
\includegraphics[width=160pt]{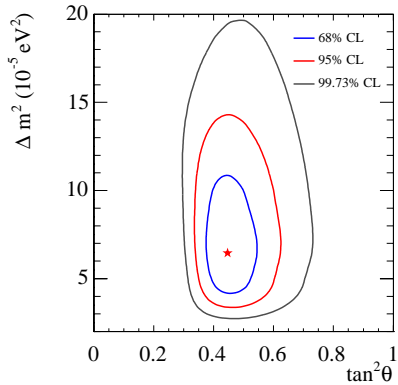}
\caption{\label{fig:plot_solar_global_nsp}  Global neutrino oscillation
analysis of solar neutrino data.  The solar neutrino data include SNO-I
day and night spectra, SNO-II day and night CC spectra and ES and NC
fluxes, the SK-I solar zenith spectra, and the rate measurements from
the $^{37}$Cl (Homestake) and $^{71}$Ga (SAGE, GALLEX, GNO)
experiments.}
\end{center}
\end{figure} 

\section{KamLAND}

Previous reactor $\bar{\nu}_e$ oscillation experiments, with
reactor-detector distances (``baselines") ranging from $\sim$10~m to
$\sim$1~km, did not observe any  $\bar{\nu}_e$ 
disappearance\cite{bib:reactor_rmp}.  If CPT is conserved and
matter-enhanced neutrino oscillation is the underlying mechanism for
the observed flavor transformation in solar neutrinos, one would expect
a significant fraction of the reactor $\bar{\nu}_e$s oscillating
(primarily through vacuum oscillation) into another flavor at a
baseline of 100 to 200~km.  The  Kamioka Liquid scintillator
Anti-Neutrino Detector (KamLAND) experiment\cite{bib:kldet} has a
unique geographic advantage over other previous reactor $\bar{\nu}_e$
experiments; it is surrounded by 53 Japanese power reactors with an
average baseline of 180~km.

KamLAND is a 1000-tonne liquid scintillator detector located in the
Kamioka mine in Japan. Its scintillator is a mixture of dodecane
(80\%),  pseudocumene (1,2,4-Trimethylbenzene, 20\%), and 1.52 g/liter
of PPO (2,5-Diphenyloxazole).  An array of 1325 17-inch-diameter PMTs
and 554 20-inch-diameter PMTs are mounted inside the spherical
containment vessel.  Outside this vessel, an additional 225
20-inch-diameter PMTs act as a cosmic-ray veto counter.

KamLAND detects $\bar{\nu}_e$s by the inverse $\beta$ decay process
$\bar{\nu}_e + p\rightarrow e^+ + n$, which has a 1.8\,MeV kinematic 
threshold.  The prompt signal $E_{\rm prompt}$ in the scintillator,
which includes the positron kinetic energy and the annihilation energy,
is related to the incident $\bar{\nu}_e$ energy $E_{\bar{\nu}_e}$ and
the average neutron recoil energy $\bar{E}_n$ by
$E_{\bar{\nu}_e}=E_{\rm{prompt}}+\bar{E}_n+0.8$~MeV.  The final state
neutron is thermalized and captured by a proton with a mean lifetime of
$\sim$200$\mu$s.  The prompt signal and the 2.2-MeV $\gamma$ emitted in
the delayed neutron capture process form a coincident signature for the
$\bar{\nu}_e$ signal.

\begin{table*}
    \caption{\label{tbl:klsummary}Summary of KamLAND $\bar{\nu}_e$ rate measurements. }
   \begin{center}
    \begin{tabular}{lll} \hline
	          & First result\cite{bib:klrate} & Second result\cite{bib:klspec}  \\ \hline\hline
	          \multicolumn{3}{c}{{\it Data Sets}} \\ \hline
Data span & Mar. 4, 2002 to Oct. 6, 2002 & Mar. 9, 2002 to Jan. 11, 2004 \\ 
Live days  & 145.1 live days & 515.1 live days \\
Exposure  & 162 ton-year & 766.3 ton-year \\ \hline \hline
		  \multicolumn{3}{c}{{\it Results}} \\ \hline
Expected signal & 86.8 $\pm$ 5.6 counts & 365.2 $\pm$ 23.7 counts \\ 
Background        & 1 $\pm$ 1 count &  17.8 $\pm$ 7.3 counts  \\ 
Observed signal &  54 counts   &  258 counts \\ 
Systematic uncertainties & 6.4\% & 6.5\% \\ 
$\bar{\nu}_e$ disappearance C.L. & 99.95\% & 99.998\% \\ \hline\hline
    \end{tabular}
    \end{center}
\end{table*}

Table~\ref{tbl:klsummary} summarizes the two KamLAND reactor
$\bar{\nu}_e$ rate measurements\cite{bib:klrate,bib:klspec} to-date.  
The exposure of the two measurements are 162~ton-years and
766~ton-years respectively.  The null hypothesis that the observed
$\bar{\nu}_e$ rates are statistical downward fluctuations is rejected
at 99.95\% and 99.998\% confidence levels in the two measurements. 
KamLAND is the first experiment that observes reactor $\bar{\nu}_e$
disappearance.

\begin{figure}
\begin{center}
\includegraphics[width=190pt]{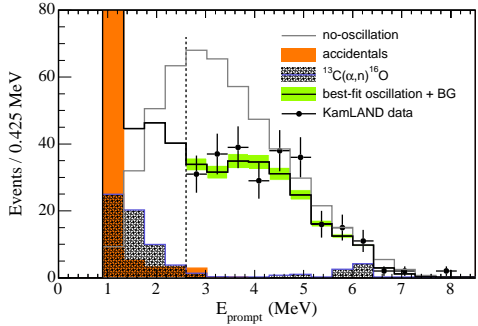}
\caption{\label{fig:kl_spect} Prompt energy spectrum of $\bar{\nu}_e$
candidate events in the 766-ton-year KamLAND analysis.  Also shown are
the no-oscillation spectrum and the best-fit spectrum under the
assumption of neutrino oscillation.}
\end{center}
\end{figure} 

\begin{figure}
\begin{center}
\includegraphics[width=160pt]{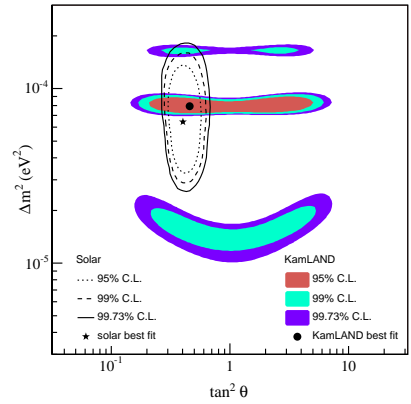}
\caption{\label{fig:kl-cont} Neutrino oscillation parameter allowed
region from the 766-ton-year KamLAND reactor $\bar{\nu}_e$ results
(shaded regions) and the LMA region derived from solar neutrino
experiments (lines).}
\end{center}
\end{figure}

If vacuum oscillation is responsible for the disappearance of reactor
$\bar{\nu}_e$s in KamLAND, then one might expect a distortion of the
$E_{\rm prompt}$ spectrum.  The measured $E_{\rm prompt}$ spectrum is
shown in Fig.~\ref{fig:kl_spect}.  A fit of the observed $E_{\rm
prompt}$ spectrum to a simple re-scaled, undistorted energy spectrum is
excluded at 99.6\%.  Also shown in the figure is the best-fit spectrum
under the assumption of neutrino oscillation.  The allowed oscillation
parameter space is shown in Fig.~\ref{fig:kl-cont}, and the best-fit
parameters of this KamLAND-only analysis are\cite{bib:klspec}
\begin{eqnarray}
	\Delta m^2 &=& 7.9 ^{+0.6}_{-0.5} \times 10^{-5} \mbox{eV}^2 \nonumber \\
	\tan^2\theta &=& 0.46. \nonumber
\end{eqnarray}

\begin{figure}
\begin{center}
\includegraphics[width=164pt]{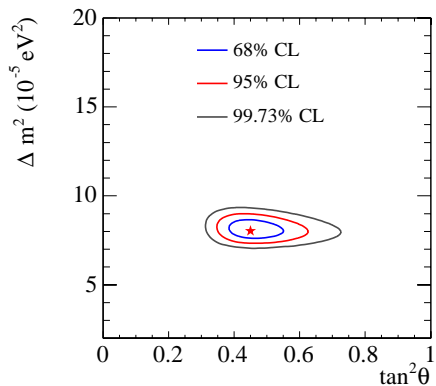}
\caption{\label{fig:global} Global neutrino oscillation analysis of
solar neutrino experiments and KamLAND.  The best fit point is at
($0.45 ^{+0.09}_{-0.07}, 8.0 ^{+0.6}_{-0.4} \times 10^{-5} \mbox{eV}^2$).}
\end{center}
\end{figure}

Assuming CPT invariance, a global neutrino oscillation analysis can be
performed on the solar neutrino results and the KamLAND results.  The
allowed parameter space is shown in Fig.~\ref{fig:global}.  The
best-fit parameters in this global analysis are\cite{bib:saltprc}
\begin{eqnarray}
	\Delta m^2 &=& 8.0 ^{+0.6}_{-0.4} \times 10^{-5} \mbox{eV}^2 \nonumber \\
	\tan^2\theta &=& 0.45 ^{+0.09}_{-0.07}. \nonumber
\end{eqnarray}
One immediately notices the complementarity of reactor anti-neutrino
and solar neutrino experiments.  The former restricts $\Delta m^2$,
whereas the latter restricts $\tan^2\theta$ in an orthogonal manner.  

\section{Future Experimental Solar Neutrino Program}

In the next several years, both running solar neutrino experiments SNO
and Super-Kamkiokande have an ambitious physics program.

The three-flavor mixing matrix element $U_{e2}$ can be
written\cite{bib:mnsp} as $\cos \theta_{13} \sin \theta_{12}$, which
approximately equals $\sin \theta$ for two-flavor solar neutrino
oscillation when $\theta_{13}$ is small and when $\Delta m^2$ from the
solar neutrino sector is much less than that from the atmospheric
neutrino sector.  For oscillation parameters in the LMA region, the MSW
effect can result in $^8$B neutrinos emerging from the Sun essentially
as a pure $\nu_2$ state.  The SNO $\phi^{CC}/\phi^{NC}$ ratio, a
direct measure of $\nu_e$ survival probability, is also a direct
measure of $\left| U_{e2} \right|^2$ ($\sim \sin^2\theta$).  Therefore,
one of the primary goals of the SNO experimental program is to make a
precision measurement of this fundamental parameter by improving on the
CC and NC measurements.

The SNO experiment has entered the third phase (SNO-III) of its physics
program. Thirty six strings of $^3$He and 4 strings of $^4$He
proportional counters have been deployed on a 1-m square grid in the
D$_2$O volume.  In SNO-I and SNO-II, the extracted CC and NC fluxes are
strongly anti-correlated (the correlation coefficient is -0.53 in
SNO-II).  This anticorrelation is a significant fraction
of the total flux uncertainties.  By introduction this array of $^3$He
proportional counters, NC neutrons are detected by $n+ ^3\mbox{He}
\rightarrow p + ^3\mbox{H}$; whereas the Cherenkov light from the CC
electrons are recorded by the PMT array.  This physical separation, as
opposed to a statistical separation of the CC and NC signals in SNO-I
and SNO-II, will allow a significant improvement in the precision of
the CC and the NC fluxes.  Table~\ref{tbl:ncdsys} summarizes the
uncertainties in the CC and NC flux measurements in SNO-I and SNO-II,
and the projected uncertainties for the corresponding measurements in
SNO-III.

Because the $^3$He proportional counter array ``removes" Cherenkov
light signals from NC interactions, it allows for a search of CC
electron spectral distortion at an analysis threshold of $T_{\rm
eff}=$4 to 4.5~MeV.  The distortion effects are enhanced at this
threshold when compared to those analyses at higher thresholds in SNO-I
and SNO-II.

\begin{table*}
    \caption{\label{tbl:ncdsys} Uncertainties in the CC and NC fluxes
    in SNO-I, SNO-II and SNO-III.  The SNO-III entries are projected
    uncertainties for 1 live year of data.  The total uncertainties are
    the quadratic sum of the statistical and systematic uncertainties.}
   \begin{center}
    \begin{tabular}{|l|cc|cc|cc|} \hline
    
        &  \multicolumn{2}{|c|}{SNO-I} & \multicolumn{2}{|c|}{SNO-II} & \multicolumn{2}{|c|}{SNO-III}  
        \\ \cline{2-7}
	& $\Delta \phi^{\rm CC}/\phi^{\rm CC}$ & $\Delta \phi^{\rm NC}/\phi^{\rm NC}$ 
	 & $\Delta \phi^{\rm CC}/\phi^{\rm CC}$ & $\Delta \phi^{\rm NC}/\phi^{\rm NC}$ 
	 & $\Delta \phi^{\rm CC}/\phi^{\rm CC}$ & $\Delta \phi^{\rm NC}/\phi^{\rm NC}$  \\ \hline

	Systematic & 5.3 & 9.0 & 4.9 & 7.3 & 3.3 & 5.2 \\ \hline
	Statistical   & 3.4 & 8.6 & 3.7 & 4.2 & 2.2 & 3.8 \\ \hline
	Total & 6.3 & 12.4 & 6.1 & 8.4 & 4.0 & 6.4 \\ \hline

    \end{tabular}
    \end{center}
\end{table*}

The Super-Kamiokande is scheduled for a detector upgrade from October
2005 to March 2006.  After this upgrade, the detector will return to
the same photocathode coverage in SK-I.  The primary physics goal of
the SK-III solar neutrino program is to search for direct evidence of
the MSW effect.  With the improvements made to the trigger system in
SK-II and the anticipated increase in photocathode coverage, the SK-III
detector will be able to push the analysis threshold down to
$E\sim$4~MeV.

\begin{figure}
\begin{center}
\includegraphics[width=190pt]{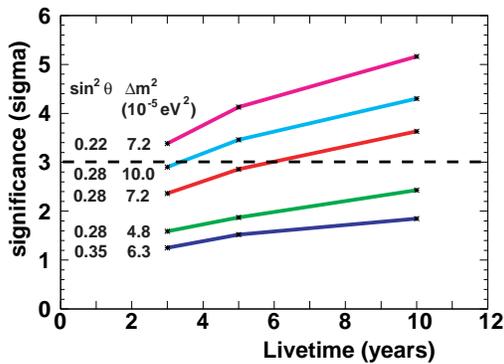}
\caption{\label{fig:sk_future-sens} Projected sensitivity to spectrum
distortion for SK-III.  Each curve represents the significance for
observing spectral distortion with the labeled mixing parameters.}
\end{center}
\end{figure}

Figure~\ref{fig:sk_future-sens} shows the projected SK-III sensitivity
to observing spectral distortion in its solar neutrino signal.  In this
figure, the sensitivities of several combinations of ($\Delta m^2$,
$\sin^2\theta$) in the  LMA region allowed by solar neutrino
measurements (c.f. Fig.~\ref{fig:plot_solar_global_nsp}) are shown.  It
is possible to discover MSW-induced spectral distortion at $>3\sigma$
after several years of counting.  

\begin{figure}
\begin{center}
\includegraphics[width=190pt]{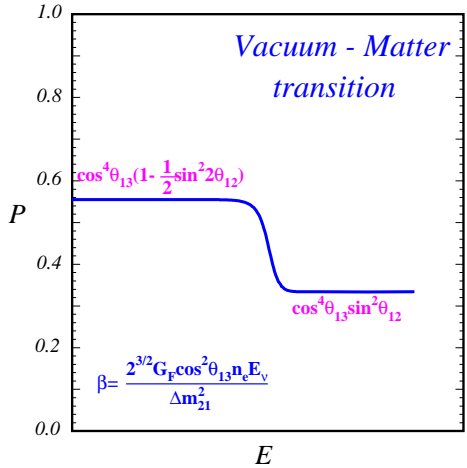}
\caption{\label{fig:vac-matter-trans} Vacuum-matter transition of 
$\nu_e$ survival probability.  In the $pp$ neutrino regime, the
survival probability is approximately $(1-\frac{1}{2}\sin^2
2\theta_{12})$ (for $\theta_{13}<<1$).  In the energy regime where
$^8$B neutrinos are observed, the survival probability is approximately
$\sin^2\theta_{12}$.  Non-standard interactions  (NSI) predict a
substantially different energy dependence in the transition region, and
future $^7$Be and the $pep$ neutrinos experiments can be used to
discriminate these NSI models from the MSW scenario.}
\end{center}
\end{figure}

Future solar neutrino experiments focus on detecting the low energy
$pp$ neutrinos ($E_\nu<0.42$~MeV), $^7$Be neutrinos ($E_\nu =
0.86$~MeV, BR=90\%), $pep$ neutrinos ($E_\nu = 1.44$~MeV), or neutrinos
from the CNO cycle.  These experiments are subdivided into two broad
classes: $\nu_e$-only detection mechanism through $\nu_e$
charged-current interaction with the target nucleus; and $\nu e$
elastic scattering which measures an admixture of $\nu_e$, $\nu_\mu$
and $\nu_\tau$.  The $pp$ neutrino experiments seek to make high
precision measurement of the $pp$ neutrino flux and to constrain the
flux of sterile neutrinos.  The $^7$Be and the $pep$ neutrino
experiments seek to map out the $\nu_e$ survival probability in the
vacuum-matter transition region.  Figure~\ref{fig:vac-matter-trans}
shows this transition region\cite{bib:bahpen}.  Although certain
non-standard interaction (NSI) models\cite{bib:nsi} can match the
survival probability in the $pp$ and the $^8$B energy regimes, they
differ substantially from the MSW prediction in the $^7$Be and $pep$
energy regimes ($E_\nu\sim 1-2$~MeV).

Table~\ref{tbl:future_sn} is a tabulation of future solar neutrino
experiments (adopted from Nakahata\cite{bib:nakahata_lowe}).  Most of
these experiments are in proposal stage; but two liquid scintillator
experiments, Borexino and KamLAND, will come  online for $^7$Be solar
neutrino measurements in the next year.  The construction of Borexino
is complete, and it is waiting for authorization to fill the detector
with liquid scintillator.  The purification system of the KamLAND
experiment is being upgraded in order to achieve an ultra-pure liquid
scintillator for the $^7$Be neutrino measurement.  

For other future solar experiments, a summary of their current status
can be found in the supplemental slides of this conference
talk\cite{bib:backup_slides}.

\begin{table*}
    \caption{ Future solar neutrino experiments.  Those experiments
    identified with an asterisk have been funded for the full scale
    detector.}
   \begin{center}
   \label{tbl:future_sn}
    \begin{tabular}{|llll|} \hline

    Experiment &  $\nu$ source & Reaction  &  Detector \\ \hline\hline
    
    \multicolumn{4}{|l|}{{\it Charged-Current Detectors}} \\ \hline
    
    LENS  &  $pp$ & $\nu_e ^{115}\mbox{In} \rightarrow e ^{115}\mbox{Sn}, e, \gamma$ &
                      15 tons of In in 200-ton liquid scintillator \\ \hline
    Lithium &  $pep$, CNO  & $\nu_e ^7\mbox{Li} \rightarrow e ^7\mbox{Be}$ &
                       Radiochemical, 10 ton lithium \\ \hline
    MOON &  $pp$ & $\nu_e ^{100}\mbox{Mo} \rightarrow e ^{100}\mbox{Tc}(\beta)$ &
    		    3.3 ton $^{100}$Mo foil + plastic scintillator \\ \hline\hline
                                       
     \multicolumn{4}{|l|}{$\nu e$ {\it Elastic Scattering Detectors}} \\ \hline                   
                       
     Borexino$^*$ & $^7$Be & $\nu e \rightarrow \nu e$ &
     		    100 ton liquid scintillator \\ \hline
     CLEAN  &  $pp$, $^7$Be & $\nu e \rightarrow \nu e$ &
     		    10 ton liquid Ne \\ \hline
     HERON &  $pp$, $^7$Be & $\nu e \rightarrow \nu e$ &
     		    10 ton liquid He \\ \hline
     KamLAND$^*$ & $^7$Be & $\nu e \rightarrow \nu e$ &
     		    1000 ton liquid scintillator \\ \hline
     SNO+  &  $pep$, CNO & $\nu e \rightarrow \nu e$ &
     		    1000 ton liquid scintillator \\ \hline
     TPC-type & 	 $pp$, $^7$Be & $\nu e \rightarrow \nu e$ &
     		    Tracking electron in gas target \\ \hline
     XMASS  & $pp$, $^7$Be & $\nu e \rightarrow \nu e$ &
     		    10 ton liquid Xe \\ \hline

    \end{tabular}
    \end{center}
\end{table*}

\section{Future Experimental Reactor Anti-neutrino Program}

The future reactor anti-neutrino program is focused on determining the
neutrino mixing angle $\theta_{13}$.  This is the only unknown angle in
the neutrino mixing matrix, and its current upper limit is
$\sin^2(2\theta_{13}) < 0.2$~(90\% C.L.)\cite{bib:chooz}.

Although there are ongoing efforts in developing accelerator-based
$\theta_{13}$ measurements by searching for $\nu_\mu \rightarrow \nu_e$
appearance, such long baseline measurements can be affected by matter
effects.  The determination of $\theta_{13}$ in these appearance
experiments is complicated by the degeneracy of mixing parameters (e.g.
$\theta_{23}$).   A reactor-based $\theta_{13}$
measurement can complement accelerator-based experiments by removing
these intrinsic ambiguities.  In a reactor-based measurement, one
searches for $\bar{\nu}_e$ flux suppression and spectral distortion of
the prompt positron signal at different baselines.

The general configuration of such a reactor-based experiment consists
of at least two or more detectors: one is placed at a distance of
$<$0.5~km, and the other at 1-2~km.  The {\it near} detector is used to
normalize the reactor $\bar{\nu}_e$ flux, while the {\it far} detector
is used to search for rate suppression and spectral distortion.  The
baseline of the far detector, $\sim$2~km, is the distance where the
survival probability reaches its first minimum.  In order to shield the
detectors from muon spallation backgrounds, these detectors require
overhead shielding. For these experiments, a $<\sim$1\% error budget is
required in order to reach a $\sin^2 2\theta_{13}$ sensitivity of
$<\sim 0.01$.

In addition to $\theta_{13}$ measurements, these experiments can also
make contributions in measuring the Weinberg angle $\sin\theta_W$ (at
$Q^2=$0), and in investigating the nature of neutrino neutral-current
weak coupling.  The precision in $\theta_{12}$ can also be improved by
a reactor anti-neutrino experiment with a baseline of 50 to
70~km\cite{bib:bandy05}.

Table~\ref{tbl:future_reactor} is a tabulation of the proposed reactor
$\theta_{13}$ experiments\cite{bib:aps_reactor}.  More details on the
status of these experiments can be found in the supplemental slides of
this talk\cite{bib:backup_slides}.

\begin{table*}
    \caption{\label{tbl:future_reactor} Proposed reactor anti-neutrino
    experiments for measuring $\theta_{13}$.}
   \begin{center}
    \begin{tabular}{|l|l|c|c|c|c|c|c|l|}  \hline
 
    &    &  \multicolumn{2}{|c|}{Baseline}  &   \multicolumn{2}{|c|}{Overburden}  &
         \multicolumn{2}{|c|}{Detector size} & $\sin^2(2\theta_{13})$ \\  
         
    Experiment & Location & \multicolumn{2}{|c|}{(km)} & 
    \multicolumn{2}{|c|}{(m.w.e.)} & \multicolumn{2}{|c|}{(tons)}  & sensitivity \\ \cline{3-8}
    
    &    &     Near & Far & Near & Far & Near & Far & (90\% C.L.) \\ \hline
    
 Angra dos Reis & Brazil & 0.3 & 1.5 & 200 & 1700 & 50 & 500 & $<\sim$0.01 \\ \hline
 
 Braidwood & USA   & 0.27 & 1.51 & 450 & 450 & 65$\times$2  & 65$\times$2 & $<\sim$0.01 \\ \hline
 
Double Chooz & France & 0.2 & 1.05 & 50 & 300 & 10 & 10 & $<\sim$0.03 \\ \hline

Daya Bay & China & 0.3 & 1.8-2.2 & 300 & 1100 & 50 & 100 & $<\sim$0.01 \\ \hline

Diablo Canyon & USA & 0.4 & 1.7 & 150 & 750 & 50 & 100 & $<\sim$0.01 \\ \hline

KASKA & Japan & 0.4 & 1.8 & 100 & 500 & 8 & 8 & $<\sim$0.02 \\ \hline

Kr2Det & Russia & 0.1 & 1.0 & 600 & 600 & 50 & 50 & $<\sim$0.03 \\ \hline

    \end{tabular}
    \end{center}
\end{table*}

\section{Conclusions}

After nearly four decades, the solar neutrino deficit problem is
finally resolved.  There are now strong evidences for neutrino
oscillation from solar neutrino and reactor anti-neutrino experiments. 
SNO, Super-Kamiokande and KamLAND will improve 
the precision of the neutrino oscillation parameters.  Future solar
neutrino experiments will provide high precision tests of the solar
model calculations, and will probe the energy dependence of neutrino
oscillation.  Future reactor anti-neutrino experiments will attempt to
measure the unknown mixing angle $\theta_{13}$.  Once the magnitude of
this angle is known, a new arena of CP studies may be opened up in the
neutrino sector in the future.

\section*{Acknowledgments}

The author thanks the conference organizers for the invitation to
speak; and E.~Blucher, M.~Chen, P.~Decowski, H.~Ejiri, R.~Hazama,
K.~Ishii, R.~Lanou, T.~Lasserre, K.-B.~Luk, D.~McKinsey, M.~Nakahata,
R.~Raghavan, F.~Suekane, R.~Svoboda, and Y.~Takeuchi for providing much
of the background information in this talk and proceedings.

 \clearpage
\twocolumn[
\section*{DISCUSSION}
]

\begin{description}
\item[Jonathan Rosner] (U. of Chicago, USA):  Can you say more about
the proposal to put liquid scintillator in SNO?

\item[Alan Poon]:  The main objectives of this proposed experiment are
to measure the {\em pep} neutrino flux and to search for geo-neutrinos.
 The {\em pep} measurement will probe the vacuum-matter transition that
I discussed in the talk.

The project is still in an early proposal stage.  There are a number of
technical problems that need to be resolved before its realisation. 
For example, the compatibility of the liquid scintillator and the
acrylic vessel has to be established; the longevity of the current
photomultiplier tube mounting structure, which was designed to have a
life span of 10 years in ultra-pure water, has to be evaluated; and the
optical response of the liquid scintillator has to be established.

\item[Peter Rosen] (DOE, USA):  A comment about MSW. If you restrict
your analysis to 2 flavors, then the fact that the solar $\nu_e$
survival probability is less than 1/2 is evidence for MSW. Pure
in-vacuum oscillation give a survival probability greater than or equal
to 1/2. Secondly, the best chance to see a day/night effect is to have
a detector at the equator. This maximizes tha path of the neutrino
through matter, and through the densest part of the Earth.

\item[Alan Poon]:  Thank you for your comment.  Because of time
constraint, I did not have enough time to discuss this point further in
my talk.  It is true that even though we have not directly observed MSW
effect, there are very strong indirect evidence that it is the
underlying mechanism for neutrino flavor transformation in  solar
neutrinos.  Fogli {\em et al.} showed that the null hypothesis of no
MSW effect in the solar neutrino results is rejected at more than
5$\sigma$.

\end{description}

\end{document}